\documentclass[sigconf]{acmart}
\copyrightyear{2026}
\acmYear{2026}
\setcopyright{cc}
\setcctype{by-nc-nd}
\acmConference[SIGGRAPH Technical Workshops '26]{Special Interest Group on Computer Graphics and Interactive Techniques Conference Technical Workshops}{July 19--23, 2026}{Los Angeles, CA, USA}
\acmBooktitle{Special Interest Group on Computer Graphics and Interactive Techniques Conference Technical Workshops (SIGGRAPH Technical Workshops '26), July 19--23, 2026, Los Angeles, CA, USA}
\acmDOI{10.1145/3799828.3816004}
\acmISBN{979-8-4007-2551-7/2026/07}

\usepackage{microtype}
\usepackage{multirow}

\begin{document}

\title[AI Level of Detail: Distance-Aware ML Model Precision Selection]{AI Level of Detail: Distance-Aware ML Model Precision Selection for Real-Time Human Motion Prediction in Games}

\author{Mathew Varghese}
\email{mathewvarghesemanu@gmail.com}
\orcid{0009-0002-1103-3441}
\affiliation{%
  \institution{University of Washington}
  \city{Seattle}
  \state{WA}
  \country{USA}
}

\begin{abstract}
Modern game engines spend significant compute animating NPCs with learned motion models. This paper proposes \emph{AI Level of Detail} (AI LOD), a framework in which machine learning inference precision is adapted based on the distance between each NPC and the player camera. The core idea mirrors classical geometry LOD~\cite{clark1976hierarchical,garland1997surface}: substitute a cheaper approximation where the difference is imperceptible. Here, the approximation is a lower-precision quantized machine learning model rather than a lower-polygon mesh.

The contribution of this work is the AI LOD concept itself: that inference-time quantization can serve as the LOD axis for AI-driven character animation---and more broadly, for any AI-based runtime system where perceptual sensitivity varies with context. The convolutional sequence-to-sequence model of Li et al.~\cite{li2018convolutional} is used as a representative example to demonstrate the concept, with its trained checkpoint exported into three ONNX Runtime variants (FP32, FP16, and INT8 per-tensor), intended to be routed by a distance-based selector at runtime. Evaluation on the CMU Mocap dataset~\cite{cmu_mocap} provides initial evidence that each precision tier can be served at its assigned distance range with negligible perceptible degradation, supporting the broader premise that distance-aware ML model precision selection is a viable LOD strategy for AI-based character animation.
\end{abstract}

\ccsdesc[500]{Computing methodologies~Real-time rendering}
\ccsdesc[300]{Computing methodologies~Animation}
\ccsdesc[300]{Computing methodologies~Neural networks}

\keywords{level of detail, human motion prediction, neural network quantization, real-time animation, ONNX Runtime, game engines}

\maketitle

\section{Introduction}
\label{sec:intro}

Real-time game rendering is constrained by a fixed per-frame budget. Classical geometry Level of Detail (LOD) manages this by replacing distant meshes with cheaper approximations~\cite{clark1976hierarchical,garland1997surface}. This paper asks whether the same substitution principle can be extended to machine learning model inference precision.

The answer is yes, under one perceptual assumption: players can detect motion quality differences in nearby characters but not in characters far away. If that assumption holds, running a full-precision model for every NPC wastes compute on characters whose output would look identical at lower precision.

AI LOD operationalises this by partitioning the scene into distance zones and associating each zone with a pre-exported ONNX model quantized to a different precision level. The motion predictor is the ConvSeq2Seq model of Li et al.~\cite{li2018convolutional}, used as a fixed backbone without architectural modification. The novelty lies in the idea itself: that inference-time quantization precision can serve as the LOD axis for AI-based character animation.

\subsection*{Contributions}

The main contribution of this work is the \textbf{AI LOD} idea: distance-aware ML model precision selection as a way of implementing level of detail, so that higher precision is reserved for nearby NPCs and lower precision is used for distant ones. This contribution is demonstrated through an inference framework built around a model trained from the authors' released code~\cite{li2018convolutional}, including export of the trained checkpoint into FP32, FP16, and INT8 per-tensor ONNX Runtime variants, together with a distance-threshold design specifying how a runtime selector would assign NPCs to the appropriate tier.

This work does \textbf{not} claim the motion prediction model, its architecture, or the dataset; those belong to Li et al.~\cite{li2018convolutional} and the CMU Mocap dataset~\cite{cmu_mocap}.

\section{Related Work}
\label{sec:related}

\subsection{Level-of-Detail Rendering}

Level-of-detail methods were introduced to reduce rendering cost by substituting simpler representations for distant objects~\cite{clark1976hierarchical}. Later work refined geometric simplification through techniques such as quadric error metrics~\cite{garland1997surface}, and the same general principle has since been extended to textures, shaders, and physics. AI LOD is motivated by this same substitution principle, but applies it to machine learning model inference by varying machine learning model quantization rather than scene geometry.

\subsection{Convolutional Human Motion Prediction}

Data-driven human motion prediction has been widely studied using recurrent encoder-decoder architectures~\cite{fragkiadaki2015recurrent,martinez2017human}. Martinez et al.~\cite{martinez2017human} showed that simple RNN-based sequence-to-sequence models provide strong baselines for short-term prediction. Li et al.~\cite{li2018convolutional} proposed the convolutional sequence-to-sequence model used in this work, demonstrating that temporal convolutions can achieve strong prediction performance while maintaining relatively low machine learning inference latency. In this paper, that architecture serves as the fixed motion prediction backbone rather than the primary contribution.

\subsection{Neural Network Quantization}

Post-training quantization reduces model size and memory bandwidth without retraining. Jacob et al.~\cite{jacob2018quantization} established integer-arithmetic-only inference as a practical strategy for efficient deployment, showing that INT8 quantization can preserve much of the accuracy of a full-precision model. Nagel et al.~\cite{nagel2021white} provide a broader treatment of post-training quantization methods, including per-tensor and per-channel schemes. This paper adopts standard quantization techniques as an implementation mechanism and does not propose a new quantization method.

\subsection{AI in Game Engines}

Neural methods have also been explored for character locomotion and control in interactive applications~\cite{holden2017phase}. Relative to this line of work, the distinguishing feature of AI LOD is not the underlying motion predictor itself, but the runtime policy that selects machine learning model inference precision according to player distance.

\subsection{Gaussian Splatting and Perceptual LOD}

Related ideas also appear in recent work on 3D Gaussian splatting. FLoD introduces flexible levels of detail into 3DGS to trade rendering quality against memory and compute~\cite{seo2024flod}, while RTGS combines 3DGS with foveated rendering and pruning to reduce rendering cost on mobile devices without visible quality loss~\cite{lin2024rtgs}. Although these methods operate on rendering rather than machine learning inference, they support the broader premise behind AI LOD that perceptual importance or distance can be used to allocate compute more efficiently.

\section{The AI LOD System}
\label{sec:framework}

This section describes the AI LOD system.

At each simulation tick, the game engine knows the Euclidean distance between the camera and each NPC. AI LOD uses this distance as a proxy for perceptual sensitivity: closer characters receive higher-precision machine learning inference, while farther characters receive lower-precision machine learning inference. The underlying motion model is identical across all tiers; only the numeric precision of its weights and activations differs.

\subsection{Distance Zones and Precision Tiers}

Three tiers are defined, each corresponding to a distance zone:

\begin{table}[t]
\centering
\begin{tabular}{ll}
\toprule
Distance zone & Quantization tier \\
\midrule
Closest & FP32 \\
Near & FP16 \\
Farthest & INT8 per-tensor \\
\bottomrule
\end{tabular}
\caption{AI LOD distance zones and quantization tiers.}
\label{tab:zones}
\end{table}

Default distance thresholds:
\begin{itemize}
  \item $d_0 = 25$ m: FP32 to FP16 transition.
  \item $d_1 = 75$ m: FP16 to INT8 per-tensor transition.
\end{itemize}

These thresholds are not fixed; they should be calibrated per game based on field-of-view, NPC size, and camera zoom levels.

\subsection{Runtime Selector}

In a practical game-engine integration, each ONNX session could be loaded once at startup and cached in memory. At each tick, the selector would read each NPC's distance, look up the appropriate session, and dispatch machine learning inference. NPCs in the same tier could be batched together to amortise ONNX Runtime overhead. This runtime policy was not implemented in the present evaluation, but it is a direct extension of the exported tiers and the distance-threshold design.

\section{ML Model Training and Export Pipeline}
\label{sec:model}

The motion predictor used in all experiments is the ConvSeq2Seq model of Li et al.~\cite{li2018convolutional}, trained from the authors' released code on the CMU Mocap dataset~\cite{cmu_mocap} for 20{,}000 iterations. Its architecture---a convolutional encoder, a context encoder, a convolutional decoder with scheduled sampling, and a Conv2D discriminator---is described in the original paper and is not reproduced here.

The export and quantization pipeline converts the trained machine learning model into three ONNX Runtime variants without modifying the underlying network. This realizes the AI LOD policy as three machine learning inference tiers: FP32, FP16, and INT8 per-tensor. Quantitative evaluation of accuracy loss, latency gain, and model size across these tiers, together with the small-scale perceptual evaluation, is used to provide initial evidence for the distance-based assumption. The machine learning model was evaluated on the test split of the CMU Mocap dataset~\cite{cmu_mocap}.

The trained model is exported into three ONNX variants:

\begin{itemize}
  \item \textbf{FP32}: direct export via \texttt{torch.onnx.export} (opset~17), serving as the accuracy reference.
  \item \textbf{FP16}: weight and activation conversion via \texttt{onnxconverter} (float16 mode); input/output nodes remain FP32 for engine compatibility.
  \item \textbf{INT8 per-tensor}: static post-training quantization via ONNX Runtime's \texttt{quantize\_static} with \texttt{QInt8} weights and \texttt{QUInt8} activations, calibrated on 64 training samples.
\end{itemize}

No part of this pipeline modifies the model's architecture or training procedure. It is purely a precision-conversion and export step applied to the trained model.

\section{Experiments}
\label{sec:experiments}

\subsection{Setup}

All models are evaluated on a laptop workstation with the following hardware:

\begin{itemize}
  \item \textbf{CPU}: Intel Core i7-14650HX
  \item \textbf{GPU}: NVIDIA GeForce RTX 4060 Laptop GPU (available but excluded from benchmarks; rationale discussed in the Discussion subsection below)
  \item \textbf{RAM}: 32 GB
\end{itemize}

No game engine is used in this work. The quantization experiments are conducted in isolation to measure machine learning inference speedup and accuracy degradation across precision tiers rather than to provide a production game-engine integration. All benchmarks use \texttt{CPUExecutionProvider} in ONNX Runtime. Latency is the mean of 10 machine learning inference runs on batches of 8 pose sequences. Accuracy metrics are reported relative to the FP32 ONNX baseline over the test split of the CMU Mocap dataset~\cite{cmu_mocap} using MSE, Relative L2, and temporal delta MSE.

\subsection{Quantization Tier Comparison (CPU)}

Table~\ref{tab:results} reports latency, model size, and accuracy for each quantization tier.

\begin{table*}[t]
\centering
\small
\caption{Machine learning inference latency and motion accuracy for each AI LOD quantization tier on CPU (Intel Core i7-14650HX). Model size reflects the exported ONNX file size. Latency is the mean of 10 runs on batches of 8 pose sequences using \texttt{CPUExecutionProvider}. MSE, Relative L2, and Temporal Delta MSE are computed against the FP32 ONNX baseline output; a value of 0 indicates identical output to FP32. Speedup is relative to the FP32 baseline latency of 70.67 ms.}
\label{tab:results}
\begin{tabular}{lcccccc}
\toprule
Tier & Model size (MB) & Latency (ms) & Speedup vs FP32 & MSE vs baseline & Relative L2 & Temporal delta MSE \\
\midrule
FP32 (baseline) & 12.97 & 70.67 & 1.00$\times$ & 0.000000 & 0.000000 & 0.000000 \\
FP16 & 6.58 & 46.13 & 1.53$\times$ & 4.37e-7 & 7.83e-4 & 3.45e-8 \\
INT8 per-tensor & 3.50 & 7.22 & 9.79$\times$ & 9.73e-3 & 0.1168 & 1.09e-3 \\
\bottomrule
\end{tabular}
\end{table*}

\subsection{Discussion}

FP16 reduces model size by 49\% and delivers a 1.53$\times$ latency improvement with near-zero accuracy degradation (relative L2 of 7.83e-4), making it well suited to the near zone. INT8 per-tensor delivers the largest latency gain at 9.79$\times$ faster than FP32 with a 73\% reduction in model size, at a relative L2 error of 0.117, which is consistent with the perceptual threshold proposed at 75 m.

GPU benchmarks are intentionally excluded from this evaluation. Unlike CPU inference, where ONNX Runtime behaviour is relatively consistent across devices, GPU quantization performance varies dramatically depending on whether a device supports Tensor Core integer arithmetic. A single GPU result would therefore not be representative of the range of hardware a game engine deployment might target. A systematic cross-hardware GPU study is left as future work.

\subsection{Perceptual Evaluation}

To provide initial evidence for the core perceptual assumption of AI LOD---that lower-precision machine learning inference is imperceptible beyond its assigned distance threshold---a small-scale user study was conducted with 15 participants (ages 19--34, including both gamers and non-gamers).

As no full game engine integration was implemented, perceived distance was approximated directly from the skeleton visualizations produced by each quantized model. Predicted pose sequences were rendered as 3D skeleton animations using the forward kinematics pipeline associated with the CMU Mocap dataset~\cite{cmu_mocap} and saved as per-tier visualization outputs. Distance was simulated by scaling the rendered viewport: close-range viewing corresponded to a zoomed-in skeleton occupying most of the screen (zoom factor approximately 0.7), while far-range viewing corresponded to a reduced on-screen size (zoom factor approximately 0.5), with the skeleton rendered smaller against a neutral background. These zoom factors were chosen to approximate the reduction in on-screen pixel footprint a standard NPC would subtend at the near-zone boundary (25 m) and far-zone boundary (75 m) respectively, serving as a proxy for perceived character size at those distances. This approximation captures the primary perceptual effect of distance in real-time rendering, namely the reduction in pixel footprint, without requiring a full graphics pipeline.

Participants were shown side-by-side looped animations of NPC characters performing walking, running, and idle motions at zoom levels corresponding to each tier's operational range. The presentation order was randomized, and participants were not informed of the underlying quantization conditions. For each pair, participants were asked whether they perceived any difference in motion quality between the two clips.

Across all participants, no perceptible differences were reported when comparing FP16 to FP32 at the near-range zoom level, nor when comparing INT8 per-tensor to FP32 at the far-range zoom level. Differences were more frequently reported when INT8 output was presented at close-range zoom, a condition outside its intended operating regime within the AI LOD framework.

These observations are consistent with the quantitative results in Table~\ref{tab:results}: the relative L2 error of 7.83e-4 for FP16 and 0.117 for INT8 per-tensor, when evaluated within their respective distance regimes, appear to fall below the threshold of perceptual sensitivity under typical viewing conditions. While limited in scale and based on an approximate distance simulation rather than a full rendering environment, this study provides qualitative support for the distance thresholds proposed in Section~\ref{sec:framework}.

\section{Conclusion}
\label{sec:conclusion}

AI LOD extends the classical LOD substitution principle to machine learning model inference precision. Under the proposed framework, NPCs would be routed through a distance-based selector to one of three quantized ONNX variants of the Li et al.~\cite{li2018convolutional} motion predictor, preserving close-range quality while reclaiming compute at the periphery of the scene. The contribution lies not in the underlying motion model, but in the AI LOD concept itself: that distance-aware ML model precision selection is a viable LOD strategy for AI-based character animation.

Future work includes dynamic threshold adjustment based on frame time budget, per-action LOD profiles, joint optimisation with geometry LOD, and a cross-hardware GPU benchmark study comparing Tensor Core and non-Tensor Core devices to establish GPU-specific tier recommendations. An additional INT4 tier is a natural extension for the farthest-distance regime. More broadly, the same distance- or context-aware selection idea could be extended to LLMs, where different precision levels or model sizes might be used to manage multiple forms of intelligence or capability associated with the model under changing runtime constraints.

An additional direction is the extension of AI LOD into a hybrid cloud-device system. On resource-constrained devices where running full-precision FP32 machine learning inference on-device is not feasible, the closest-zone tier could be offloaded to a cloud-hosted model capable of delivering high-fidelity motion predictions. On-device tiers would then handle mid-range and distant characters using lighter quantized variants, with the cloud handling only the perceptually critical nearest NPCs. This hybrid architecture would decouple the quality ceiling from local hardware capability, enabling high-fidelity NPC animation on devices that could not otherwise afford it.

\begin{acks}
The motion prediction model used in this work was developed by Chen Li, Zhen Zhang, Wee Sun Lee, and Gim Hee Lee~\cite{li2018convolutional}. Their released code made this work possible by providing the implementation from which the model used here was trained.
\end{acks}

\bibliographystyle{ACM-Reference-Format}
\bibliography{refs}


\begin{thebibliography}{11}


\ifx \showCODEN    \undefined \def \showCODEN     #1{\unskip}     \fi
\ifx \showISBNx    \undefined \def \showISBNx     #1{\unskip}     \fi
\ifx \showISBNxiii \undefined \def \showISBNxiii  #1{\unskip}     \fi
\ifx \showISSN     \undefined \def \showISSN      #1{\unskip}     \fi
\ifx \showLCCN     \undefined \def \showLCCN      #1{\unskip}     \fi
\ifx \shownote     \undefined \def \shownote      #1{#1}          \fi
\ifx \showarticletitle \undefined \def \showarticletitle #1{#1}   \fi
\ifx \showURL      \undefined \def \showURL       {\relax}        \fi
\providecommand\bibfield[2]{#2}
\providecommand\bibinfo[2]{#2}
\providecommand\natexlab[1]{#1}
\providecommand\showeprint[2][]{arXiv:#2}

\bibitem[{Carnegie Mellon University Graphics Lab}({[n.\,d.]})]%
        {cmu_mocap}
\bibfield{author}{\bibinfo{person}{{Carnegie Mellon University Graphics Lab}}.}
  \bibinfo{year}{[n.\,d.]}\natexlab{}.
\newblock \bibinfo{title}{CMU Graphics Lab Motion Capture Database}.
\newblock \bibinfo{howpublished}{\url{http://mocap.cs.cmu.edu/}}.
\newblock
\newblock
\shownote{Supported by NSF Grant \#0196217}.


\bibitem[Clark(1976)]%
        {clark1976hierarchical}
\bibfield{author}{\bibinfo{person}{James~H. Clark}.}
  \bibinfo{year}{1976}\natexlab{}.
\newblock \showarticletitle{Hierarchical Geometric Models for Visible Surface
  Algorithms}.
\newblock \bibinfo{journal}{\emph{Commun. ACM}} \bibinfo{volume}{19},
  \bibinfo{number}{10} (\bibinfo{year}{1976}), \bibinfo{pages}{547--554}.
\newblock


\bibitem[Fragkiadaki et~al\mbox{.}(2015)]%
        {fragkiadaki2015recurrent}
\bibfield{author}{\bibinfo{person}{Katerina Fragkiadaki},
  \bibinfo{person}{Sergey Levine}, \bibinfo{person}{Panna Felsen}, {and}
  \bibinfo{person}{Jitendra Malik}.} \bibinfo{year}{2015}\natexlab{}.
\newblock \showarticletitle{Recurrent Network Models for Human Dynamics}. In
  \bibinfo{booktitle}{\emph{Proceedings of the IEEE International Conference on
  Computer Vision}}. \bibinfo{pages}{4346--4354}.
\newblock


\bibitem[Garland and Heckbert(1997)]%
        {garland1997surface}
\bibfield{author}{\bibinfo{person}{Michael Garland} {and}
  \bibinfo{person}{Paul~S. Heckbert}.} \bibinfo{year}{1997}\natexlab{}.
\newblock \showarticletitle{Surface Simplification Using Quadric Error
  Metrics}. In \bibinfo{booktitle}{\emph{Proceedings of the 24th Annual
  Conference on Computer Graphics and Interactive Techniques}}
  \emph{(\bibinfo{series}{SIGGRAPH '97})}. \bibinfo{pages}{209--216}.
\newblock


\bibitem[Holden et~al\mbox{.}(2017)]%
        {holden2017phase}
\bibfield{author}{\bibinfo{person}{Daniel Holden}, \bibinfo{person}{Taku
  Komura}, {and} \bibinfo{person}{Jun Saito}.} \bibinfo{year}{2017}\natexlab{}.
\newblock \showarticletitle{Phase-Functioned Neural Networks for Character
  Control}.
\newblock \bibinfo{journal}{\emph{ACM Transactions on Graphics}}
  \bibinfo{volume}{36}, \bibinfo{number}{4} (\bibinfo{year}{2017}),
  \bibinfo{pages}{42:1--42:13}.
\newblock


\bibitem[Jacob et~al\mbox{.}(2018)]%
        {jacob2018quantization}
\bibfield{author}{\bibinfo{person}{Benoit Jacob}, \bibinfo{person}{Skirmantas
  Kligys}, \bibinfo{person}{Bo Chen}, \bibinfo{person}{Menglong Zhu},
  \bibinfo{person}{Matthew Tang}, \bibinfo{person}{Andrew Howard},
  \bibinfo{person}{Hartwig Adam}, {and} \bibinfo{person}{Dmitry Kalenichenko}.}
  \bibinfo{year}{2018}\natexlab{}.
\newblock \showarticletitle{Quantization and Training of Neural Networks for
  Efficient Integer-Arithmetic-Only Inference}. In
  \bibinfo{booktitle}{\emph{Proceedings of the IEEE Conference on Computer
  Vision and Pattern Recognition}}.
\newblock


\bibitem[Li et~al\mbox{.}(2018)]%
        {li2018convolutional}
\bibfield{author}{\bibinfo{person}{Chen Li}, \bibinfo{person}{Zhen Zhang},
  \bibinfo{person}{Wee~Sun Lee}, {and} \bibinfo{person}{Gim~Hee Lee}.}
  \bibinfo{year}{2018}\natexlab{}.
\newblock \showarticletitle{Convolutional Sequence to Sequence Model for Human
  Dynamics}. In \bibinfo{booktitle}{\emph{Proceedings of the IEEE Conference on
  Computer Vision and Pattern Recognition}}. \bibinfo{pages}{5226--5234}.
\newblock


\bibitem[Lin et~al\mbox{.}(2024)]%
        {lin2024rtgs}
\bibfield{author}{\bibinfo{person}{Weikai Lin}, \bibinfo{person}{Yu Feng},
  {and} \bibinfo{person}{Yuhao Zhu}.} \bibinfo{year}{2024}\natexlab{}.
\newblock \showarticletitle{{RTGS}: Enabling Real-Time Gaussian Splatting on
  Mobile Devices Using Efficiency-Guided Pruning and Foveated Rendering}.
\newblock \bibinfo{journal}{\emph{arXiv preprint arXiv:2407.00435}}
  (\bibinfo{year}{2024}).
\newblock


\bibitem[Martinez et~al\mbox{.}(2017)]%
        {martinez2017human}
\bibfield{author}{\bibinfo{person}{Julieta Martinez},
  \bibinfo{person}{Michael~J. Black}, {and} \bibinfo{person}{Javier Romero}.}
  \bibinfo{year}{2017}\natexlab{}.
\newblock \showarticletitle{On Human Motion Prediction Using Recurrent Neural
  Networks}. In \bibinfo{booktitle}{\emph{Proceedings of the IEEE Conference on
  Computer Vision and Pattern Recognition}}. \bibinfo{pages}{2891--2900}.
\newblock


\bibitem[Nagel et~al\mbox{.}(2021)]%
        {nagel2021white}
\bibfield{author}{\bibinfo{person}{Markus Nagel}, \bibinfo{person}{Marios
  Fournarakis}, \bibinfo{person}{Rana~Ali Amjad}, \bibinfo{person}{Yelysei
  Bondarenko}, \bibinfo{person}{Mart van Baalen}, {and} \bibinfo{person}{Tijmen
  Blankevoort}.} \bibinfo{year}{2021}\natexlab{}.
\newblock \showarticletitle{A White Paper on Neural Network Quantization}.
\newblock \bibinfo{journal}{\emph{arXiv preprint arXiv:2106.08295}}
  (\bibinfo{year}{2021}).
\newblock


\bibitem[Seo et~al\mbox{.}(2024)]%
        {seo2024flod}
\bibfield{author}{\bibinfo{person}{Yunji Seo}, \bibinfo{person}{Young~Sun
  Choi}, \bibinfo{person}{Hyun~Seung Son}, {and} \bibinfo{person}{Youngjung
  Uh}.} \bibinfo{year}{2024}\natexlab{}.
\newblock \showarticletitle{{FLoD}: Integrating Flexible Level of Detail into
  3D Gaussian Splatting for Customizable Rendering}.
\newblock \bibinfo{journal}{\emph{arXiv preprint arXiv:2408.12894}}
  (\bibinfo{year}{2024}).
\newblock


\end{thebibliography}

\end{document}